\documentclass[fleqn,twoside]{article}
\usepackage{espcrc2,epsfig}

\usepackage{graphicx}
\usepackage[figuresright]{rotating}

\newcommand{\AmS}{{\protect\the\textfont2
  A\kern-.1667em\lower.5ex\hbox{M}\kern-.125emS}}

\title{Electric Dipole Moments of Fundamental Particles}

\author{Yannis K. Semertzidis\address[BNL]{Brookhaven National Laboratory, 
        Physics Department, \\ 
        P.O. Box 5000, Upton, NY 11973-5000, USA}%
}
       
\def\ecm{\rm e \cdot cm}

\begin{document}

\begin{abstract}

Electric dipole moment (EDM) experiments are at the fore-front of search for physics 
beyond the standard model.
The next generation searches promise to improve by several orders
of magnitude the current EDM sensitivity levels. 
\vspace{1pc}
\end{abstract}

\maketitle

\section{Theoretical Motivation}

The search for electric dipole moments (EDM) of fundamental 
particles started approximately
fifty years with Ramsey's search for a neutron EDM.  Even though
the techniques have been improved and the sensitivities has reached
an unprecedented small level no EDM of a fundamental particle has
been observed so far.  Non-the-less EDM experiments have put strict limits and 
constrained the parameter space of models beyond the standard model.

The permanent EDM of fundamental particles would violate both time 
(T) and parity (P) symmetries:  an EDM vector would have to be along
the spin vector since there is no other defining vector. 
Phenomenologically any component in any other direction would 
average out to zero due to the particle's spin rotation.
The interaction Hamiltonian is given
by

\begin{equation}
H_E = -d_E \vec{S} \, \cdot \vec{E}
\end{equation}
where $\vec{S}, \, \vec{E}$ denote the spin vector and the electric field 
vector respectively. The symbol $d_E$ denotes  the electric dipole moment
strength.  Under
parity the axial vector $\vec{S}$ does not change sign whereas $\vec{E}$
does.  The opposite happens when the time operator is applied, i.e. the
vector $\vec{S}$ does change sign whereas $\vec{E}$
does not.  In both cases the interaction Hamiltonian changes sign meaning 
that if $d_E$ is not zero the Hamiltonian
would violate both parity and time reversal symmetries.

This is not the case for the magnetic dipole moments since the interaction 
Hamiltonian in that case is

\begin{equation}
H_M = -d_M \vec{S} \, \cdot \vec{B}
\end{equation}
Under parity both axial vectors $\vec{S}$ and $\vec{B}$ do not change
sign whereas under time reversal they both do. Therefore the interaction
Hamiltonian does not suffer a sign change and the parity and time
reversal symmetries are respected by the magnetic dipole moments.

If EDMs are not allowed by the above symmetry considerations how then
there can be induced EDMs, permanent EDMs of polar molecules, etc?
 The cases of the induced electric dipole moments 
are allowed since the EDM vector in those cases is proportional to the electric
field vector $\vec{d}_E= \alpha \vec{E}$ and not the spin vector.  The 
interaction Hamiltonian becomes proportional to the square of the E-field
and both symmetries, parity and time reversal, are respected.  As far as the
polar molecules that exhibit ``permanent'' electric dipole moments they also respect
the above symmetries with their quantum mechanical treatment  described by
Penny~\cite{Penny31}.

Through the fundamental CPT conservation theorem, T-violation also means 
CP-violation.  A general overview of the importance of CP-violation is written
by J. Ellis in the CERN Courier~\cite{Ellis99} in October of 1999.
Sakharov~\cite{Sakharov67} in his 1967 paper pointed out that
CP-violation is one of three requirements needed to explain the matter antimatter
asymmetry of our universe.     

The first requirement was that the proton
 should be unstable. The second was that there would be 
interactions violating C and CP and the 
third condition was that the universe would 
undergo a phase of extremely rapid expansion.

\subsection{EDMs are Excellent Probes of Physics Beyond the SM}

\begin{figure}[htb]
\vspace{9pt}
\epsfig{file=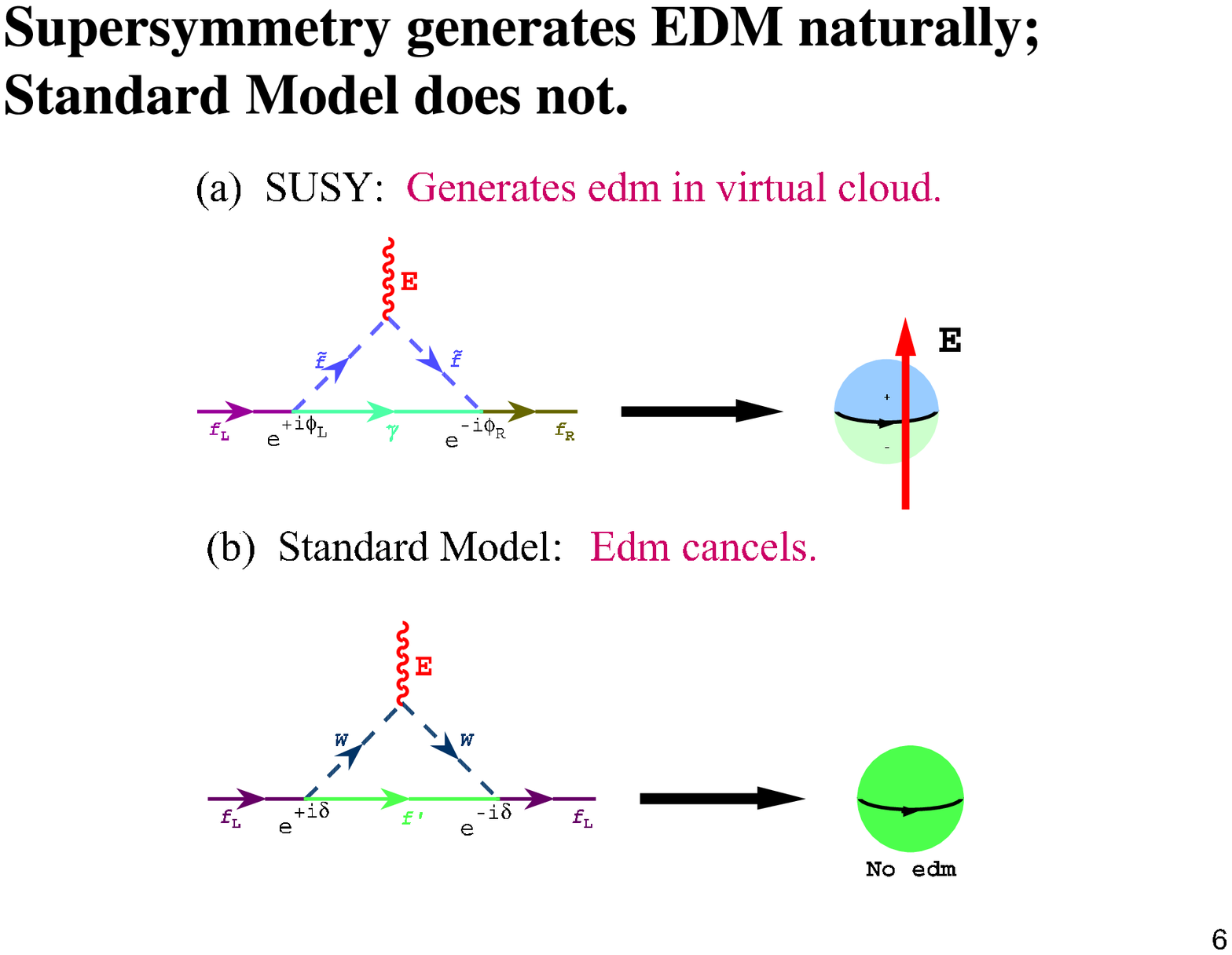,width=6.5cm,angle=-90}
\vskip -1 cm
\caption{In many models, like SUSY, EDMs are non-zero at the one loop level  
 but the SM EDMs are zero at that level. This is so because there is only 
one CP-violating phase in the SM, and the $W$ boson only couples to left handed
particles.  In contrast SUSY has more than 40 CP-violating phases, plus sfermions 
couple to both left and right-handed particles making unnatural the
first order cancellation of EDMs. The figure is copied from
reference~\cite{fortson03}.}
\label{fig:susy_sm}
\end{figure}

In the standard model (SM) there is only one CP-violating phase
(KM) which results to an EDM only after third order loops with
 virtual $W^{\pm}$s and  quarks
are considered. This results to a natural
suppression of the SM EDMs by several orders of magnitude.  
In contrary, physics models beyond the SM allow for much higher values
of EDM, see Figure~\ref{fig:susy_sm} (from ref.~\cite{fortson03}), 
many times in the experimentally accessible region.  
For example super-symmetry (SUSY) has more than 40
CP-violating phases and the first order EDM calculation does not
cancel as it does in the SM.  Other models with similar EDM predictions
include models with left-right symmetry, multi Higgs scenarios, etc.

\section{Experimental Approach}

The spin of a particle with an electric dipole moment $d$ precesses in the
presence of an electric field.  Since the $d$ value is presumably 
very small (non observed so far) the spin precession signal would be of very
small frequency.  A magnetic field is used to serve as a carrier signal by
pressing the spin due to its magnetic dipole moment.  The spin precession
rate is given by

\begin{equation}
{d\vec{S} \over dt} = \vec{\mu} \times \vec{B} + \vec{d} \times \vec{E}
\end{equation}
For a spin 1/2 particle ${d\vec{S} \over dt} ={1 \over 2} \hbar \omega$, 
where $\omega$ is called the Larmor frequency.  In case of an atomic or 
molecular electron the magnetic field causes a spectral split in the 
line and the transitional frequency is called Zeeman splitting.
One then compares the Larmor/Zeeman frequencies with the E-field vector  flipped
back and forth:
$\hbar (\omega_1 - \omega_2) = 4 d E$.  In order to reduce the effects of
a drifting magnetic field another particle with an expected small
EDM sensitivity value is used as a B-field sensor, also known as co-magnetometer.

\subsection{Schiff theorem}

The experimental approach was influenced by the 
Ramsey-Purcell-Schiff theorem~\cite{ramsey} which
 states that for point like, charged 
particles in equilibrium the net electric field they feel averages to zero.
In an external electric field the electronic and nuclear charge
 of an atom would be re-arranged
so that the net (average) electric field on all charged particles
 would be zero,
known as ``Schiff's theorem''.  Otherwise
they would be continuously accelerated.  However as was pointed out by Schiff himself
and others~\cite{khriplovich}
 not all the forces need to
be electrical.  The electric field can thus be compensated by magnetic,
nuclear, etc. forces and even though the total force is zero
there is a net electrical force. This results to a non-zero EDM value for the
atom or molecule, called ``Schiff's moment''. Sandars further pointed out that in
paramagnetic atoms,
there is even an enhancement of the average electric field the unpaired electron feels
in the presence of an external electric field when relativistic effects are 
taken into account.  The reason for the enhancement is due to
the very strong electric fields present near the nucleus.  The enhancement factor
calculated by Sandars~\cite{sandars} is given by 
$R= d_a/d_e \approx 10 Z^3 \alpha^2$ which for
large size atoms can be quite a big factor. $d_a$, is the atomic electric dipole moment
and $d_e$ that of the electron, $Z$ is the atomic number and $\alpha$ the fine 
structure constant.  As an example $R=115$ for the Cs atom and $R=-585$ for the
Tl atom.  Sandars work is the basis so far of all the searches for the electron EDM
with atoms or molecules.

\subsection{Electron EDM}

The current experimental electron EDM limit comes from the Berkeley atomic thallium
experiment~\cite{commins02}.  It is a small scale, ``table top'', experiment where 
Tl atomic beams are led to go through high electric field regions where
there is also a magnetic field present.  The Larmor frequency is probed with the
standard technique of Ramsey separated fields.  Motional magnetic fields of the
form $\vec{u} \times \vec{B}$, with $\vec{u}$ the atomic beam velocity can be
a problem in the presence of small misalignments between the $\vec{E}$ and $\vec{B}$
fields.  Another potential systematic error is Berry's phase and some 8 atomic 
beams with fluxes over $10^{18}$ atoms/sec are used to study the systematic effects using
many different correlations.  The final result of this experiment is 
$|d_e| < 1.6 \times 10^{-27} \ecm$ (90\% C.L.)~\cite{commins02}.

\subsection{$\rm ^{199}Hg$ EDM}

\begin{figure}[htb]
\vspace{9pt}
\epsfig{file=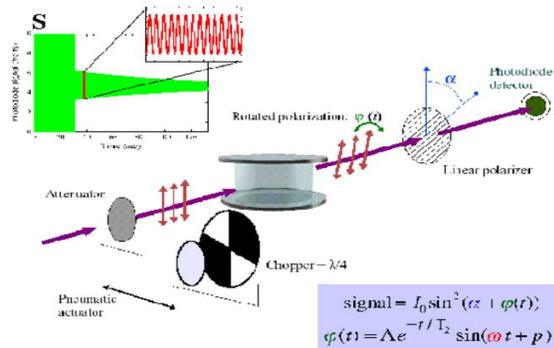,width=8cm,height=5.5cm}
\vskip -1. cm
\caption{The experimental principle of the $^{199}Hg$ experiment; from 
reference~\cite{fortson03}.}
\label{fig:hg_method}
\end{figure}

The mercury EDM experiment is a ``table top'' effort at Washington state.~\cite{romalis}
They look for a shift in the Zeeman frequency in $^{199}Hg$ vapor when the E-field is 
flipped. The mercury vapor is contained in two adjacent vapor cells where the B and
E-fields are parallel.  The mercury atoms are polarized by circularly polarized laser 
light of 254~nm modulated (chopped) at the Larmor frequency.  A plane (linearly)
 polarized laser goes through the cell with the mercury vapor where its plane
of polarization rotates according to $\alpha \approx \vec{k} \cdot \vec{S} $, with $\vec{k}$
the laser propagation vector and $\vec{S}$  the mercury spin, precessing in the horizontal
plane at the Larmor frequency, Figure~\ref{fig:hg_method}.
 The statistical accuracy of the method is given by

\begin{equation}
\delta d = {\hbar \over 2 E \sqrt{N\tau T}}
\end{equation}

with $N$ the number of observed photons, $E$ the electric field strength, $\tau$ the spin
coherence time and $T$ the total running time of the experiment.  The result is 
$|d(^{199}Hg)| < 2.1 \times 10^{-27} \ecm$, (90\% C.L.)~\cite{romalis,fortson03}.

\vskip -1 cm

\subsection{Neutron EDM}

The neutron EDM experiments have first started 50 years ago and have come along way
since. Currently ultra cold neutrons (UCN) from a nuclear reactor are brought to
a region where a large electric field is present along with a small magnetic field.
The neutron EDM principle also uses the technique of Ramsey separated fields to probe the
Larmor precession.  A potential EDM signal is any phase shift correlated with the
reversal of the electric field vector. 
In order to minimize systematic errors due to the magnetic field drifts  atomic
$^{199}Hg$ is used as a co-magnetometer.  The combined results give a limit of
$|d_n|<6.3 \times 10^{-26} \ecm$ 
(90\% C.L.)~\cite{romalis2}.  An upgrade of the Grenoble experiment where more neutrons
are to be collected and has as a goal  $|d_n|<1.5 \times 10^{-26} \ecm$~\cite{grinten}.

\subsection{Prospects}

  The next generation experiments, are very promising.  
On the electron 
a Yale group under D. DeMille made great progress towards using the metastable
molecule of PbO$^*$~\cite{kawall}.  The group promises an order of magnitude improvement
over the current electron limits within a year 
and another two orders within the next couple
of years.
S. Lamoreaux has described~\cite{lamoreaux} a solid state technique where the
alignment of the atoms in an electric  field would align the spins of the atoms and hence it
will lead to the magnetization of the sample. This work is well underway at
Los Alamos and it promises to reach $10^{-31} \ecm$ or so within a year or two.

\vskip -1. cm

\begin{figure}[htb]
\vspace{9pt}
\epsfig{file=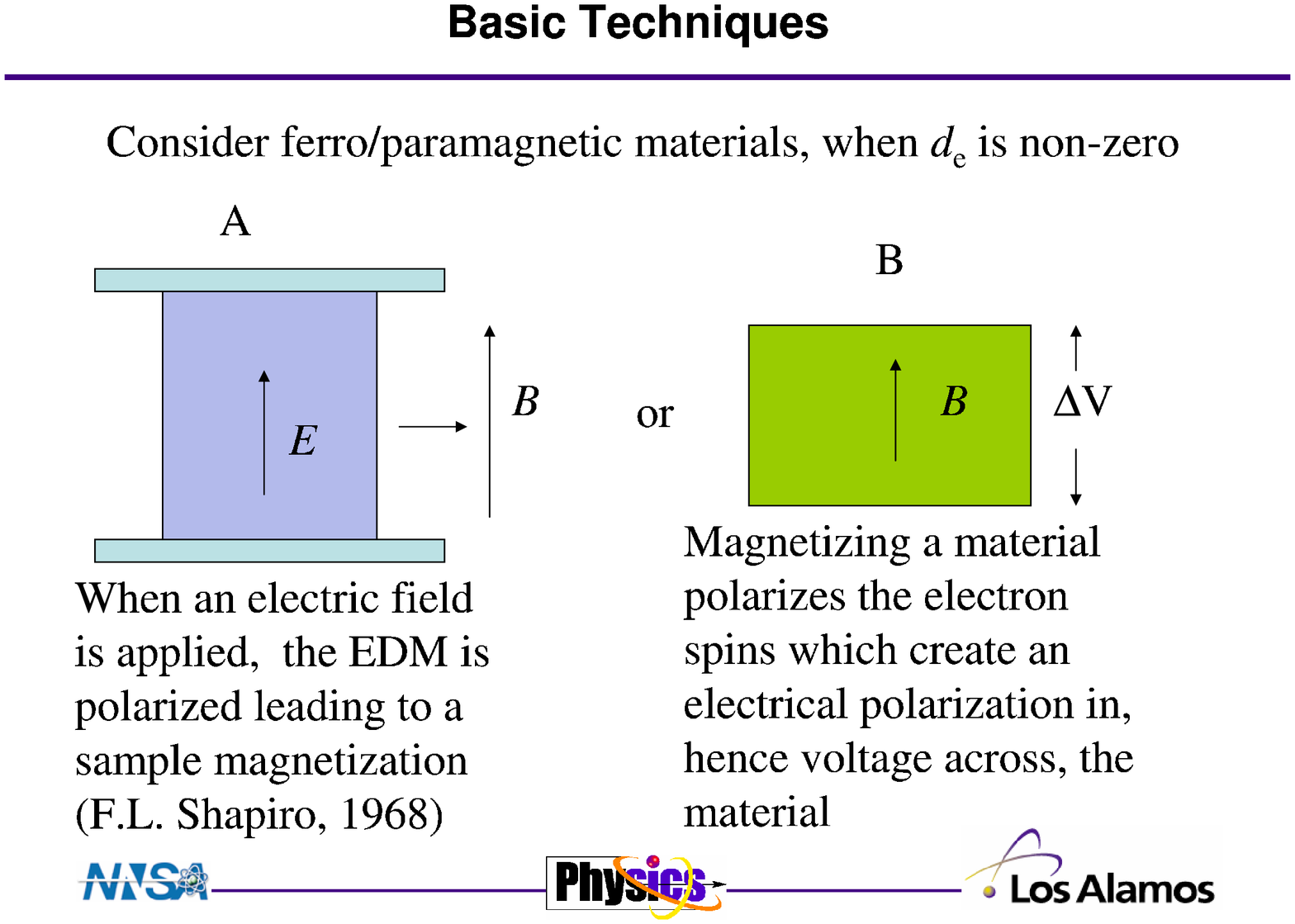,width=5cm,angle=-90}
\vskip -0.5 cm
\caption{A schematic diagram of the new neutron EDM experiment experiment of Los Alamos. From 
reference~\cite{lamoreaux}.}
\label{fig:pol}
\end{figure}

On the $^{199}Hg$ EDM experiment the Washington group has upgraded the experiment,
using four cells, the middle two with opposite electric field and the outer two
without any electric field present, in order to monitor the magnetic field 
fluctuations.  An improvement of the order of a factor of four is expected
by the group when the experiment is done~\cite{fortson03}.

The Los Alamos neutron EDM effort uses a very high flux of UCN in superfluid
$^4He$.  Polarized $^3He$ is used to probe the neutron spin precession and as a
co-magnetometer.  The neutron spin precession is probed by the reaction 
$^3 \rm \vec {H}e + \vec n \rightarrow t +p$
the cross section of which is $<10^2$~b when their spins are parallel 
and $\approx 10^4$~b when the spins
are opposite.  Since the gyromagnetic ratio of $^3He$ is within 10\% the same
as the neutron's the beat signal frequency is 10 times smaller making it 10 times
less sensitive
to the magnetic field fluctuations.  One of the challenges of the experiment is
to avoid even a single spark of the 50~kV/cm electric field since that would
surely destroy the SQUID system that is needed to monitor the magnetic field.

\vskip -1. cm

\begin{figure}[htb]
\vspace{9pt}
\epsfig{file=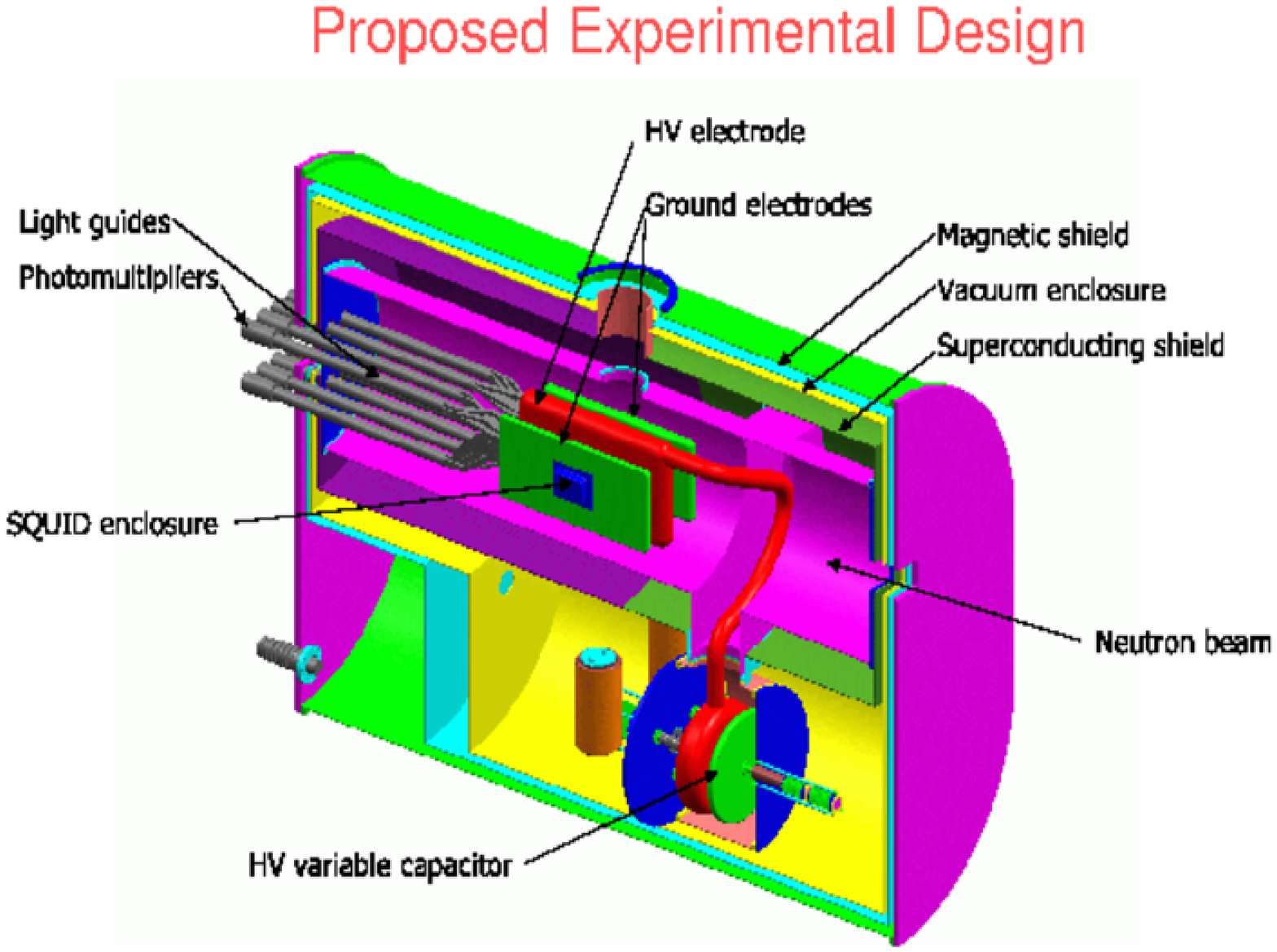,width=6cm,angle=-90}
\vskip -0.1 cm
\caption{A schematic diagram of the new neutron EDM experiment experiment of Los Alamos.
From reference~\cite{lamoreaux}.}
\label{fig:neutron2}
\end{figure}

Clearly the competition between the different
EDM experiments is intense (it's a horse race!)
 and their prospects of finding a non-zero EDM value
are very good.

\section{EDM in Storage Rings}

Other than using either atoms, molecules or neutrons in the search of EDM it is possible
to look for EDM of charged particles in storage rings. The Schiff theorem does not hold
here since the particles are in an accelerated frame.  One, in principle,
does not need an electric field present in the lab frame to probe the EDM of the particle.
Even in a purely magnetic field storage ring there is an electric field in the particle's
rest frame due to Lorentz transformation: $\vec{E} = \gamma (\vec{u} \times \vec{B})$, 
with $\vec{u}$ the particle's velocity.  Since the average $\vec{B}$ vector is vertical
the induced electric field is radial and the spin, due to an EDM, 
will precess in the vertical direction.

Compared to the traditional methods of searching for EDM one notices that the electric
and magnetic fields that the particle feels in its own rest frame are strongly coupled
and one cannot significantly change their values independently.  Furthermore there is no
way of flipping the sign of the electric field while keeping the magnetic field sign the 
same. 
More over, since the electric and magnetic fields in the particle's rest frame are not
parallel but orthogonal to each other, the EDM effect on the particle is to precess its
spin in a plane orthogonal to the g-2 precession plane.\cite{CERN}
 As a result the EDM effect is a small disturbance on the regular g-2 precession.

Non-the-less this method of searching for EDM was used by the CERN as well as the 
BNL g-2 experiments.\cite{CERN,bennett:2002mg}  A new, dedicated method of searching for
``EDM in Storage Rings'' has been developed~\cite{edm_prl,himus99} in which the
g-2 precession vector is cancelled by a radial electric field.  
A major development
in this method was the realization that the EDM signal sign changes between clockwise (CW)
and counter-clockwise (CCW) storage.\cite{edm_prl} This method  has regained the 
advantages of a traditional EDM search and works
best for charged 
particles with small anomalous magnetic moment values
like the muon and deuteron promising several
orders of magnitude sensitivity improvement over current methods.

\subsection{Muon EDM}

It also happens that the same particles provide new opportunities:  The muon is the only second
generation particle that can be probed at a very sensitive level.  Furthermore it is the 
only system that can be probed in its elementary form and not as part of another system.
Therefore its interpretation would be more straightforward than any other system.

The muon anomalous magnetic moment, $a_\mu$, and electric dipole moment, $d_\mu$,
can be related to each 
other~\cite{Mar_99,Graesser:2001ec,Feng:2001sq} as the real and imaginary parts of a more general
dipole moment, D.  

\begin{eqnarray}
a_\mu {e \over 2 m_\mu} &=& \Re D \label{amu0}\\
d_\mu                   &=& \Im D \label{dmu0},
\end{eqnarray}
where $ \Re D$ and $ \Im D$ are correspondingly the real and imaginary parts of $D$.
Writing $D^{NP} = |D^{NP}| \, e^{i \phi_{\rm{CP}} }$ as the contribution of ``New Physics'' to $D$ 
provides a measure of the relative probing power of $a_\mu$ and  $d_\mu$ experiments.
If ``New Physics'' gives rise to a discrepancy between experiment and Standard Model
expectations, $a_\mu^{\rm NP} = a_\mu^{\rm exp} - a_\mu^{\rm SM}$, then one expects that
same ``New Physics'' to induce a muon EDM given by

\begin{equation}
d_{\mu} \simeq 3 \times 10^{-22} \, \left( \frac{a_\mu^{\rm NP}}{3 \times 10^{-9}} \right)
\, \tan\phi_{\rm{CP}} \, ~\ecm\  .
\label{phicp0}
\end{equation}
Of course, the values of $a_\mu^{\rm NP}$ and $\tan\phi_{\rm{CP}}$ are model dependent.

For the current situation (assuming the $e^+ e^-$ data for the hadronic 
contribution)~\cite{bennett:2004,davier2,jeg}

\begin{equation}
a_\mu^{\rm exp} - a_\mu^{\rm SM} \simeq 3(1) \times 10^{-9} \label{difference}
\end{equation}
one expects

\begin{equation}
d_{\mu} \simeq 3 \times 10^{-22} \, \tan\phi_{\rm{CP}} \, ~\ecm\  .
\label{phicp1}
\end{equation}
So, exploring down to $d_\mu \sim 10^{-24} \, \ecm$ would probe

\begin{equation}
\tan\phi_{\rm{CP}} \geq 3(1) \times 10^{-3}.
\end{equation}

Within specific models, predictions for the muon EDM vary 
widely~\cite{pilaftsis,babu,matchev,ellis,strumia}. 
In particular, the left-right supersymmetric model with the seesaw
mechanism of reference~\cite{Babu:2000dq}
predicts      $d_{\mu}$  as large as $5 \times 10^{-23}~\ecm$, 50 times larger than
the sensitivity of the proposed experiment.  The prediction for the EDM of the electron
is of order $10^{-28}~\ecm$, 10 times smaller than the present experimental 
limit~\cite{commins02}.

\subsection{Muon EDM Experimental Approach}

In the presence of both electric and magnetic fields, oriented 
orthogonally to the muon velocity and to each other,
the angular frequency of muon spin precession relative to the momentum is given by

\begin{equation}
\vec{\omega}  = {e \over m} \left\{a \vec{B}  + \left({1 \over{\gamma^2 -1}}
- a \right)
{{\vec{\beta} \times \vec{E} } \over c} + 
 {\eta \over 2} \left({\vec{E} \over c} + \vec{\beta} \times \vec{B}\right)
\right\} , \label{eq:spin2}
\end{equation}
where $a=(g-2)/2$ and $\eta$ is the EDM in units of 
${e\hbar \over 4m c}$. 

  The magnetic and electric dipole
moments are given by 
$\mu = {g\over 2} {e \hbar \over 2 m }$ and
$d   = {\eta \over 2} {e \hbar \over 2 m c}$, respectively. 
$\eta$ plays a role for the EDM corresponding to  
the g factor for the magnetic dipole moment.
The muon EDM couples to the
external fields through the $\eta(\vec E+ c \vec \beta \times\vec B)$ term.
The external B-field couples to the EDM because it produces an
E-field in the rest frame of the muon.
In fact, for the parameters envisioned in the present proposal, the
motional E-field from the $\vec \beta \times\vec B$ term is far larger
than that due to the applied E-field.
The EDM value
is given in terms of the dimensionless parameter $\eta$ by

\begin{equation}
d_{\mu} = {\eta  \over 2} {e \hbar \over 2 m_\mu c} \simeq \eta \times 4.7
\times
10^{-14} \, {\rm e \cdot cm}.         \label{eq:edmmu}
\end{equation}
for the muon.

Assuming that the EDM is 0, from  Eq.~(\ref{eq:spin2}), 
it is clear that  at the ``magic'' $\gamma$, 
($\gamma = 29.3$) 
\begin{equation}
{1 \over{\gamma^2 -1}} - a = 0 , \label{eq:magic}
\end{equation}

\noindent{and} the muon spin precession depends only on g-2 and 
the average B-field. 
The anomalous precession frequency, due to the {\it magnetic}
moment, is measured by observing the time 
spectrum of muon decay electrons. In the muon rest frame, the
highest energy electrons are emitted preferentially along the muon
spin vector. As the spin vector precesses relative to the momentum
vector, the number of high energy electrons observed in the lab
frame is modulated at the precession frequency.

\begin{figure}[htb]
\vspace{9pt}
\epsfig{file=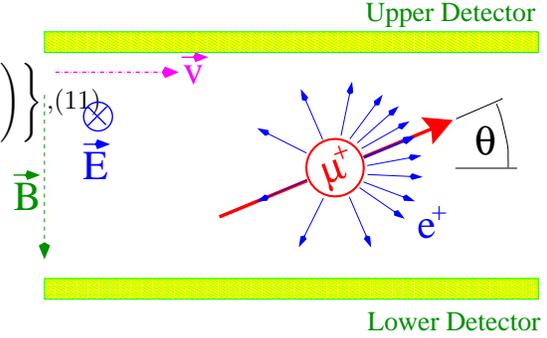,width=4.5cm,angle=-90}
\vskip -0.7 cm
\caption{ Due to the spatial anisotropy in the 
decay $\mu^+ \rightarrow e^+ 
         + \nu_e + \overline{\nu_{\mu}}$ detectors above and below the
storage region are expected 
         to observe a time dependent change in the ratio of positron
counting signals. The  
         positron angular distribution is indicated by the density of
arrows. }
\label{fig:principle}
\end{figure}

 For the dedicated EDM experiment proposed in this document we will
follow a new approach:
 Use muons with much lower energies, and
employ a radial electric field which
cancels the g-2  precession.
 The electric field in the lab required to cancel the g-2 
precession is

\begin{equation}
E \simeq a B c \beta \gamma^2,
 \label{eq:efield}
\end{equation}
 which we will  assume here equal  to about 2~MV/m.
 Using Eqs.~(\ref{eq:spin2}, \ref{eq:efield}) the  spin precession
angular frequency is given by:

\begin{equation}
\vec{\omega}  = -{e \over m} 
 {\eta \over 2} \left({\vec{E} \over c} + \vec{\beta} \times \vec{B}\right), \label{eq:spin3}
\end{equation}
 i.e. the g-2 precession is canceled and only the EDM is left to act
on the spin. The torque in the center of mass is given by

\begin{equation}
{d\vec S/dt'} =
{\vec d \times \vec E'}.
\label{eq:edm_spin}
\end{equation}

\noindent
which in terms of laboratory quantities is
 
\begin{equation}
{d\vec S/dt} =
{\vec d \times (\vec E + c \vec \beta \times \vec B)}.
\label{eq:edm_spin2}
\end{equation}
 As previously mentioned,
for realizable values for the applied E-field, the ``motional'' E-field
from the $\vec \beta \times \vec B$ term is much larger than that from
the $\vec E$ term.

\begin{figure}[htbp]
\begin{center}
\epsfig{file=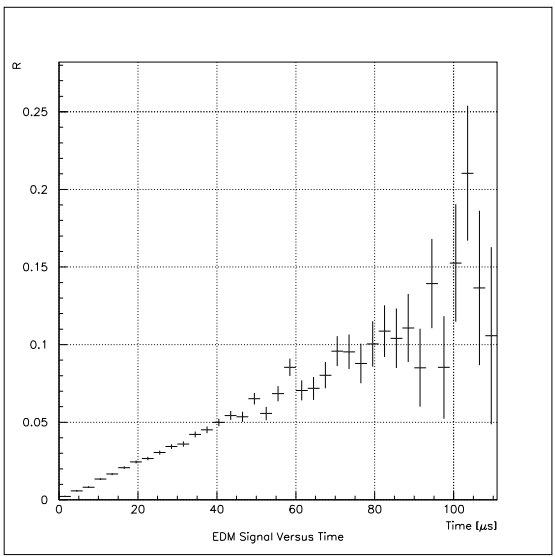,width=6cm,angle=0}
\end{center}
\hfill
\vskip -0.3 cm
\caption[signal]{{MC simulation of the muon EDM signal, 
$R={N_{up}-N_{down}\over N_{up}+N_{down}}$, versus time.}
\label{fg:signal}}
\end{figure}

   Thus the muon spin direction will  be ``frozen'' relative to
the muon momentum if the EDM is zero.  In the presence of a non-zero EDM,
the radial E-field in the muon's rest frame 
will cause rotation of the spin in a vertical plane about
an axis parallel to the radial direction.
As the spin acquires a vertical component,
the decay positron momenta  also acquire a vertical component,
resulting in an up-down asymmetry in the number,
$R_N={N_{up}-N_{down}\over N_{up}+N_{down}}$
of electrons 
which   grows linearly with time,
see Figs.~(\ref{fig:principle},\ref{fg:signal}). 
Together with other improvements, which will significantly
reduce many systematic errors, this new experimental approach will
improve our sensitivity to 
a muon EDM by five orders of magnitude.

\subsection{Deuteron EDM}

The situation with the deuteron is similar to the muon with the difference that
the deuteron 
does not decay and can, in principle, be stored for a long time. The limitation to the EDM
measurement is the spin coherence time $\tau_p$, i.e. the time the deuteron beam is stored without
losing its polarization.
 Another difference
is that it is much heavier than the muon, its spin is 1, and its anomalous magnetic moment is 
$a = -0.143$.  The fact that its spin is 1 it means that it has both a vector and a
tensor polarization which can complicate its detection.  The fact that its anomalous magnetic
moment is negative means that the electric field direction needs to be radially outward,
in opposite direction than the muon case.

The statistical accuracy of the deuteron experiment is estimated to be

\begin{equation}
\sigma_d \approx 6.5 {\hbar a \gamma^2 \over \sqrt{\tau_p} E_R (1 + a \gamma^2) A P \sqrt{N_c f T_{Tot}}}
\end{equation}
where $\tau_p \approx 10 s$ is the spin coherence time of the stored beam,
$A \approx 0.3$ is the left/right asymmetry observed by the polarimeter~\cite{polarimeter}
 when the deuteron beam is completely vertically polarized, 
$P \approx 0.55$ the polarization of the beam,
$N_c \approx 10^{11}$~d/cycle
 the total number of stored particles per cycle, $f \approx 0.01$ the useful event rate fraction, and
$T_{Tot} \approx 10^7$~s the total running time of the experiment, and $E_R \approx 3.5$~MV/m the radial
electric field strength. Then $\sigma_d \approx 5 \times 10^{-28} \ecm$ but it is estimated~\cite{cipanp_yan}
that due to the presence of the tensor polarization there will be a total loss in running time due
to the need to run for systematic error determinations of about a factor of 16, or a factor of 4 in
statistical error, i.e. $\sigma_d \approx 2 \times 10^{-27} \ecm$.

The current status of the deuteron EDM effort is that the collaboration is considering writing
a proposal to do this experiment with the above sensitivity.  There are three candidate places to
host it:  Brookhaven National Lab, Groningen University-KVI in The Netherlands, and
Indiana University Cyclotron Facility.

A deuteron EDM at the  $ 10^{-27} \ecm$ level constitutes an improvement in the sensitivity of 
the T-odd nuclear forces by a factor of 100 over the $^{199}Hg$ EDM experiment, a factor of
100,000 improvement over the current proton EDM limit and a factor of 50-100 over the current neutron
EDM experiment~\cite{khriplovich2}.


\vskip 1cm


\begin{thebibliography}{9}
\bibitem{Penny31} W.G. Penny, Phil. Mag. {\bf 11}, 602 (1931).
\bibitem{Ellis99}$\rm www.cerncourier.com/main/article/39/8/16$, October 1999.
\bibitem{Sakharov67} ``Violation of CP Invariance, C Asymmetry, and Baryon
Asymmetry of the Universe'',
 A.D. Sakharov, 1967. 
Reprinted in Kolb, E.W. (ed.), Turner, M.S. (ed.): 
The early universe 371-373, and in 
Lindley, D. (ed.) et al.: Cosmology and particle physics
 106-109, and in Sov. Phys. Usp. 34 (1991) 392-393 
[Usp. Fiz. Nauk 161 (1991) No. 5 61-64]. 
Published in Pisma Zh.Eksp.Teor.Fiz.5:32-35,1967, 
JETP Lett.5:24-27,1967, Sov.Phys.Usp.34:392-393,1991, Usp.Fiz.Nauk 
161:61-64,1991 (No.5).
\bibitem{fortson03} Fortson's talk at Lepton-Moments, Cape Cod, 9-12 June 2003, 
$\rm http://g2pc1.bu.edu/~leptonmom/program.html$.
\bibitem{ramsey} L.I. Schiff, Phys. Rev. {\bf 132}, 2194 (1963).
\bibitem{khriplovich} V.F. Dmitriev, I.B. Khriplovich, and V.B. Telitzin,
Phys. Rev. {\bf C50}, 2358 (1994) and references therein.
\bibitem{sandars} P.G.H. Sandars, Phys. Lett. {\bf 14}, 194 (1965); 
Phys. Lett. {\bf 22}, 290 (1966).

\bibitem{commins02} B.C. Regan {\it et al.,} ``New Limit on the Electron Electric Dipole Moment'',
Phys.\ Rev.\ Lett.\  {\bf 88}, 071805 (2002).

\bibitem{kawall} D. Kawall {\it et al.}, hep-ex/0309079 
\bibitem{romalis}	M.V. Romalis {\it et al.}, Phys. Rev. Lett. {\bf 86}, 2505 (2001).
\bibitem{romalis2} M.V. Romalis, ICAP 2003 proceedings.
\bibitem{grinten} V. der Grinten,  talk at Lepton-Moments, Cape Cod, 9-12 June 2003. 
\bibitem{lamoreaux} S. Lamoreaux,  talk at Lepton-Moments, Cape Cod, 9-12 June 2003. 
\bibitem{CERN}J. Bailey, K. Borer, F. Combley, H. Drumm, F.J.M. Farley, 
J.H. Field, W. Flegel, P.M. Hatterley, F. Krienen, F. Lange, E. Picasso, and
W. von R\"{u}den, J. Phys. {\bf G4}, 345 (1978);  J. Bailey {\it et al.}, Nucl. Phys. {\bf B150}, 1
(1979).

\bibitem{bennett:2002mg} G.~W.~Bennett {\it et al.}  
[Muon g-2 Collaboration],
``Measurement of the positive muon anomalous magnetic moment to 0.7~ppm,''
Phys.\ Rev.\ Lett.\  {\bf 89}, 101804 (2002)
[hep-ex/0208001].

\bibitem{edm_prl}F.J.M. Farley et al., hep-ex/0307006, submitted to PRL.
\bibitem{himus99} Y.K. Semertzidis et al., hep-ph/0012087, Proceedings of HIMUS99 Workshop,
Tsukuba, Japan (1999).
\bibitem{Mar_99} 
W. Marciano, HIMUS99 Workshop, Tsukuba, Japan (1999).

\bibitem{Graesser:2001ec} M.~Graesser and S.~Thomas,
``Supersymmetric relations among electromagnetic dipole operators,''
hep-ph/0104254.


\bibitem{Feng:2001sq} 
J.~L.~Feng, K.~T.~Matchev and Y.~Shadmi,
``Theoretical expectations for the muon's electric dipole moment,''
hep-ph/0107182.


\bibitem{bennett:2004} G.~W.~Bennett {\it et al.}  
[Muon g-2 Collaboration],
``Measurement of the negative muon anomalous magnetic moment to 0.7~ppm,''
submitted Phys.\ Rev.\ Lett.\   (2004).

\bibitem{davier2}M. Davier, S. Eidelman, A. H\"ocker, Z. Zhang  Aug 2003,
Eur. Phys. J. {\bf C 31}, 503 (2003).


\bibitem{jeg} S. Ghozzi and F. Jegerlehner, hep-ph/0310181,
  Phys. Lett. {\bf B}, in Press, (2004).



\bibitem{pilaftsis} A. Pilaftsis, Nucl. Phys. {\bf B644},
263 (2002).
\bibitem{babu}  K.S. Babu,
B. Dutta, and R.N. Mohapatra, Phys. Rev. Lett. {\bf 85}, 5064 (2000).
\bibitem{matchev}J.L. Feng, K.T. Matchev, and Yael Shadmi, Nucl. Phys. {\bf B613},
366 (2001).
\bibitem{ellis} J.R. Ellis et al., Phys. Lett. {\bf B528}, 86 (2002).
\bibitem{strumia} A. Romanino and A. Strumia, Nucl. Phys. {\bf B622}, 73 (2002); B. Dutta
and R.N. Mohapatra, Phys. Rev. {\bf D68}, 113008 (2003); A. Bartl et al., 
Phys. Rev. {\bf D68} 053005 (2003); T. Feng et al., Phys. Rev. {\bf D68}, 016004 (2003);
I. Masina, Nucl. Phys. {\bf B671} 432 (2003); G.C. Branco and D. Delepine,
Phys. Lett. {\bf B567}, 207 (2003);  I. Masina, Nucl. Phys. {\bf B661} 365 (2003).


\bibitem{Babu:2000dq}
K.~S.~Babu, B.~Dutta and R.~N.~Mohapatra,
``Enhanced electric dipole moment of the muon in the presence of large  neutrino mixing,''
Phys.\ Rev.\ Lett.\  {\bf 85}, 5064 (2000)
[hep-ph/0006329].

\bibitem{polarimeter}    L.M.C. Dutton, et al., Phys. Lett. {\bf 16}, 331 (1965);
    L.M.C. Dutton, et al., Phys. Lett. {\bf B25}, 245  (1967);
    K.S. Chadha and V.S. Varma, Phys. Rev. {\bf C13}, 715 (1976);
    L.M.C. Dutton, et al., Nucl. Phys. {\bf A343}, 356 (1980);
    B. Bonin, et al., NIM {\bf A288}, 389 (1990);
    V.P. Ladygin, et al., NIM {\bf A404}, 129 (1998);
    J. Arvieux, et al., NIM {\bf A273}, 48 (1988).

\bibitem{cipanp_yan}Y.K. Semertzidis et al., hep-ex/0308063, CIPANP proceedings (2003).
\bibitem{khriplovich2}	V.V. Flambaum, I.B. Khriplovich, and O.P. Sushkov, Phys. Lett. B162 (1985) 213;
	I.B. Khriplovich and R.A. Korkin, Nucl. Phys. A665 (2000) 365; M Pospelov et al.,
private communication (2003).

\end{thebibliography}
\end{document}